# Cluster dynamics modeling of materials: advantages and limitations.

A. Barbu and E. Clouet

Service de Recherche de Métallurgie Physique, CEA/Saclay 91191 Gif-sur Yvette, France

alain.barbu@cea.fr

**Abstract.** The aim of this paper is to give a short review on cluster dynamics modeling in the field of atoms and point defects clustering in materials. It is shown that this method, due to its low computer cost, can handle long term evolution that cannot, in many cases, be obtained by Lattice Kinetic Monte Carlo methods. Indeed, such a possibility is achieved thanks to an important drawback that is the loss of space correlations of the elements of the microstructures. Some examples, in the field of precipitation and irradiation of metallic materials are given. The limitations and difficulties of this method are also discussed. Unsurprisingly, it is shown that it goes in a very satisfactory way when the objects are distributed homogeneously. Conversely, the source term describing the primary damage under irradiation, by nature heterogeneous in space and time, is tricky to introduce especially when displacement cascades are produced.

**Introduction.**

In material science, cluster dynamics (CD) is based on kinetic equations describing the formation and evolution of clusters of solute atoms or point defects such as vacancies or self interstitial atoms (SIA). It is a very efficient method in term of computational cost. This efficiency is due to a drawback coming from the basic hypothesis of uniform distributions of clusters: the real system is replaced by an effective medium in which all processes occur continuously in time and space. The spatial correlations between clusters is consequently not considered explicitly. Lattice Kinetic Monte Carlo methods do not suffer of such a limitation. However, they are very time consuming compared with the CD meso-scale modeling and, are even unable to address long time aging in many cases. Consequently, CD remains at this time an irreplaceable method for predicting long term evolution of materials. In a first part, the CD modeling is presented. In the second one, examples of applications to precipitation and clustering of point defects under irradiation are given.

**Cluster dynamics models**

In these models, the system is seen as a gas of clusters made of monomers that can be solute atoms, vacancies, self interstitial atoms. A consequence is that the total volume fraction occupied by the clusters must be small. The evolution of the number density of clusters of each size $n$ and of a certain kind is treated within the framework of the chemical rate theory. In the most general case, i.e. when some species are non-conservative, it is described by a set of differential equations of the form:

$$\frac{dC_n}{dt} = \sum_m J_{m \to n} - \sum_q J_{n \to q} + G_n - K_n C_n \quad (1)$$

where $C_n$ is the cluster density of size $n$ and $J_{m \to n}$ the cluster flux from the class of size $m$ to the class $n$. It is given by:

$$J_{m \to n} = \sum_m w_{m \to n} C_m \quad (2)$$

with $w_{m \to n}$, the transition rate per unit concentration from the class of size $m$ to the class of size $n$.

$G_n$, the production rate of clusters of size $n$ and $K_n$, the loss of clusters of size $n$ at some fix sinks such as dislocations, grain boundaries or surfaces, must only be taken into account in the case of non conservative species as, vacancies or SIAs.

Putting (2) in (1) we have:

$$\frac{dC_n}{dt} = G_n + \sum_m w_{m \to n} C_m - \sum_q w_{n \to q} C_n - K_n C_n$$

(3)

The cluster distributions can only be calculated numerically. Keeping in mind that the interest of CD is its short computer time, the maximum size of the cluster that can be considered is limited by the number of differential equations.

A detailed description, as the one given by eq.3, is important for small cluster sizes. On the contrary, an approximated description can be used for large sizes. For this purpose, two approaches have been considered:

The first one, introduced by Ghonien et al [1], consists in developing eq.3 to the second order about $n$. Omitting $G_n$ and $K_n$, the system evolution is described by the Fokker Planck type equation:

$$\frac{\partial C(x)}{dt} = \frac{\partial}{\partial x}[f(x)C(x)] + \frac{\partial^2}{\partial x^2}[d(x)C(x)]$$

(4)

where $C(x)$ is the size distribution function. $f(x)$ and $d(x)$ are functions of the transition rates. The size of the cluster, $x$, is now a continuous variable. This continuous equation can be numerically solved discretizing the continuous variable $x$, with $\Delta x$ increasing with $x$. This method is simple but has the disadvantage of being unstable if the numerical scheme is not adequate and does not strictly conserve the matter. The instabilities can be avoided by using a numerical scheme insuring the positiveness of the concentrations. Concerning the matter losses, the discretization must be chosen to minimize it. In any case, the losses must be checked afterward to know if they are acceptable.

The second one is the grouping method consisting in replacing a group of master equations by only one equation representing the class [2]. It has been shown that the result can be very bad if the grouping is not carried out properly. Furthermore, as the previous one, it does not conserve strictly the matter even with an optimized grouping. Recently, Golubov [3] proposed a new grouping method that can conserve at the same time the matter and the cluster number. For this purpose, only the first and the second moment of the distribution are considered. Two equations for each class i.e. twice more are to be considered. The equations on the first moment ensure the conservation of the number of precipitates and those on the second moment, the conservation of matter.

**Application to precipitation.**

Let us consider a solid solution of B atom in a matrix A. Assuming that, (i) only single B atoms are mobile, (ii) only single B atoms can be emitted from a precipitate, (iii) precipitates have a well defined concentration and shape in such a way that they are only characterized by one number $n_B$, the number of B atoms they contain, the flux of clusters is given by:

$$J_{n_B \to n_B+1} = \beta_{n_B} C_{1_B} C_{n_B} - \alpha_{n_B+1} C_{n_B+1}$$

(5)

Where $C_{n_B}$ is the number of clusters per unit volume and made of n B atoms, $\beta_{n_B}$ is the condensation rate and $\alpha_{n_B}$, the emission rate of B atoms. If the condensation is controlled by diffusion, the expression of $\beta_{n_B}$, generally used is:

$$\beta_{n_B} = 4\pi r_{n_B} D_{1_B}$$
(6)

where $r_{n_B}$ is the capture radius of a cluster containing $n_B$, B atoms, $C_{1_B}$ and $D_{1_B}$ the concentration and diffusion coefficient of B atoms. Eq.6 is in principal only valid for an isolated precipitate in an infinite media. Brailford and al [4] have given a $\beta_{n_B}$ self-consistent expression showing a number density and radius precipitate dependence (muli-sink correction). However, eq.6 is usually valid as far as the volume fraction of precipitate is not very high.

To calculate the emission rate, it is classically assumed that it is an intrinsic property of the cluster and consequently that it does not depend on the concentration in the solid solution. So, it is the same as the one calculated for a solid solution at equilibrium for which $J_{n_B+1 \to n_B}$ is equal to zero. The evaporation rate is then given by:

$$\alpha_{n_B} = 4\pi r_{n_B-1} \frac{D_1}{\Omega} \exp\left[(G_{n_B} - G_{n_B-1} - G_1)/kT\right]$$
(7)

Where $G_{n_B}$ is the free energy of a cluster containing $n_B$ B atoms and, $\Omega$, the atomic volume.

In the framework of the capillary model, currently used in the classical theory of nucleation, and assuming that the atomic volume of solute atoms is the same as the solvent atom one, (no size effect contribution), $G_{n_B}$ is given by:

$$G_{n_B} = n_B \left[\mu_B(C_B^0) - \mu_A(C_B^0)\right] + \frac{n_B}{c_p} \Delta G^{nuc}(C_B^0) + \left(\frac{36\pi\Omega^2}{c_p^2}\right)^{1/3} n_B^{2/3} \sigma_{n_B}$$
(8)

Where $\Delta G^{nuc}(C_B^0)$ is the nucleation free energy of a solution of concentration $C_B^0$ and $\sigma_{n_B}$ the interface free energy of a cluster containing $n_B$, B atoms. This yield:

$$\alpha_{n_B} = 4\pi r_{n_B-1} \frac{D_1}{\Omega} \exp\left[\left(\frac{36\pi\Omega^2}{c_p^2}\right)^{1/3} \left\{n_B^{2/3} \sigma_{n_B} - (n_B-1)\sigma_{n_B-1} - \sigma_{1_B}\right\}/kT\right]$$
(9)

Contrary to the classical theory of nucleation, the nucleation free energy $\Delta G^{nuc}(C_B^0)$ does not appear in the expression of the rate of emission in CD modeling. The reason is that the gas of cluster is by itself a thermodynamic model of the alloy [5].

A typical evolution of the size histograms of Al$_3$Sc precipitates calculated with parameters representative of the AlSc0.18at% solid solution is given on figure 1 [6].

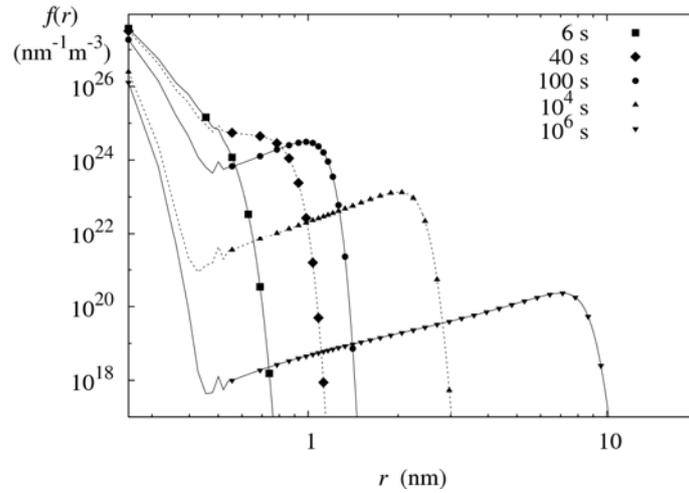

**Figure 1:** Size histogram evolution of Al$_3$Sc precipitates in the AlSc0.18at% solid solution thermally treated at 350°C.

A comparison between CD and KMC results, calculated with the same parameters, shows that CD gives a good extrapolation of the mean radius evolution obtain with KMC at time that cannot be reached by KMC. Further, it is clear that only CD simulations allow comparisons with experiments that, in this case, are very good.

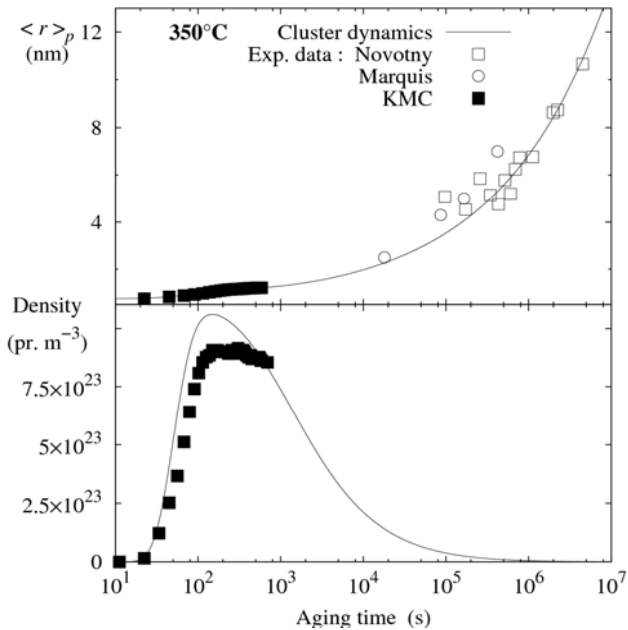

Figure 2: Mean radius and number density of precipitates as a function of the aging time for the Al Sc 0.18 at%Sc solid solution aged at 350°C as given by cluster dynamics and by Kinetic Monte Carlo. Experimental results are also reported [7-8].

It is worth noticing that the surface energy of the precipitates $\sigma_{n_B}$, used in the CD calculations to have a good agreement with KMC must depend on $n_B$. It is given by a law of the form: $\sigma_{n_B} = \bar{\sigma}\left(1 + c\, n_B^{-1/3} + d\, n_B^{-2/3}\right)$ where $\bar{\sigma}$ and the $c$ and $d$ coefficients are deduced from the Kinetic Monte Carlo results [6]. Furthermore, the discrete values must be used for $n_B<10$.

**Application to microstructural evolution under irradiation.**

The main difficulty with the clustering of point defects under irradiation is that all the objects we have to deal with are not conservative: interstitial and vacancy type objects are produced during the collisions of irradiating particles with the atoms of the material. They can also recombine or disappear on fix sinks (dislocations, surfaces or grain boundaries). It is generally considered that, at least in very pure metals, the interstitial and vacancy clusters are mobile (in alloys, the point defect clusters are likely less mobile). The master equation for interstitial and vacancy cluster evolutions are consequently more complex. Assuming for simplicity that only single vacancy and SIAs are mobile, it is, for interstitial clusters such as n>2, given by:

$$\frac{dC_{ni}}{dt} = G_{ni} + \left(\beta^i_{(n-1)i} C_{1i}\right) C_{(n-1)i} + \left(\beta^v_{(n+1)i} C_{1v} + \alpha^i_{(n+1)i}\right) C_{(n+1)i}$$
$$- \left(\alpha^i_{ni} + \beta^v_{ni} C_{1v} + \beta^i_{ni} C_{1i}\right) C_{ni} \quad (10)$$

where:

$C_{ni}$ is the concentration of SIA clusters containing $n$ SIAs.

$G_{ni}$ is the production rate of SIA clusters.

$\beta^i_{ni} C_{1i}$ is the agglomeration rate of an SIA with a SIA cluster of $n$ SIAs, producing an interstitial cluster of $n+1$ SIAs.

$\beta^v_{ni} C_{1i}$ is the reaction rate of a vacancy with a SIA cluster of $n$ SIAs, leading to an interstitial cluster of $n-1$ SIAs.

$\alpha^i_{ni}$ the emission rate of a SIA from a interstitial cluster of $n$ SIAs.

$\alpha^v_{ni}$ the emission rate of a single vacancy from an interstitial cluster of $n$ SIAs.

A symmetrical equation with symmetrical definitions yields for vacancy clusters.

The equation form is significantly different for the mobile species, here the single vacancies and SIAs. For SIAs (symmetrical equation for vacancies), it is given by:

$$\frac{dC_{1i}}{dt} = G_{1i} + \beta^v_{2i} C_{1v} C_{2i} - C_{1i} \sum_{m=1} \beta^i_{mi} C_{mi} - \beta^i_{1i} C^2_{1i} - C_{1i} \sum_{m=1} \beta^i_{mv} C_{mv} - \beta^v_{1i} C_{1i} C_{1v}$$
$$+ \sum_{m=2} \alpha^i_{mi} C_{mi} + \alpha^i_{2i} C_{2i} + \sum_{m=2} \alpha^i_{mv} C_{mv} \quad (11)$$

Indeed, the mobile species can react with all clusters whatever their type (third and fifth term). Further, clusters of two monomers and more can in principal emit a single defect of the same or of the opposite type (seventh and ninth terms). The probability for a vacancy cluster to emit an SIA is in reality so low that this term can be erased. The reverse is not true: the probability for an interstitial dislocation loop to emit a vacancy cannot be ignored if the size of the loop and the temperature are large enough. The reason of the fourth term is that when a monomer of one type encounter a monomer of the same type, two monomers disappear and for the eighth term, that when a dimmer dissociates, two monomers appears.

Note that in these equations, the classical recombination term between a vacancy and a SIA is splitted in two terms: the first term of the second sum and the sixth term.

The agglomeration coefficient is given by a relation of the same type as eq.6 [9-12]. However, the effective capture radius depends in the shape of the cluster (dislocation loops for interstitial clusters and dislocation loops, voids or staking fault tetrahedras for vacancy clusters). The interaction between point defects and the point defect clusters must be considered. It is particularly true for SIA and dislocation loops whatever their nature.

The emission coefficients are given by a relation of the same form as eq.7 [9-12]. In the case of emission of a SIA from an SIA cluster, it can be rewritten:

$$\alpha^i_{ni} = 4\pi r^i_{(n-1)i} \frac{D_{1i}}{\Omega} \exp\left[E^{Bi}_{ni} / kT\right] \quad (12)$$

Where $r^i_{ni}$ is the spherical capture radius of a SIA by a SIA cluster containing $n$ SIAs and $E^{Bi}_{ni}$ the binding energy of a SIA with a SIA cluster containing $n$ SIAs. The same kind of relation holds for the other emitting coefficients. As for precipitation, the $E^{Bi}_{ni}$s must be calculated at the atomic level for small clusters. In order to limit the number of parameters $E^{Bi}_{ni}$, extrapolation can be used for large clusters. The extrapolation laws have usually the same form as the capillary law used for precipitation of solute atoms [11].

The reaction rate for the elimination of defects on surfaces or grain boundaries has been given by Bullough et al [13]. In this case, contrary to point defect clusters, the multi-sink effect can never be omitted as, it appears in the first order term.

The large number of parameters entering in this kind of modeling is often an object of criticisms. They are on our opinion non relevant. Indeed, all parameters correspond to a physical reality. Either they are assumed not to be important and, in this case, CD modeling is a good method to check it, or they are important and, in this case, the models ignoring them for simplification, does not correspond to the physical reality.

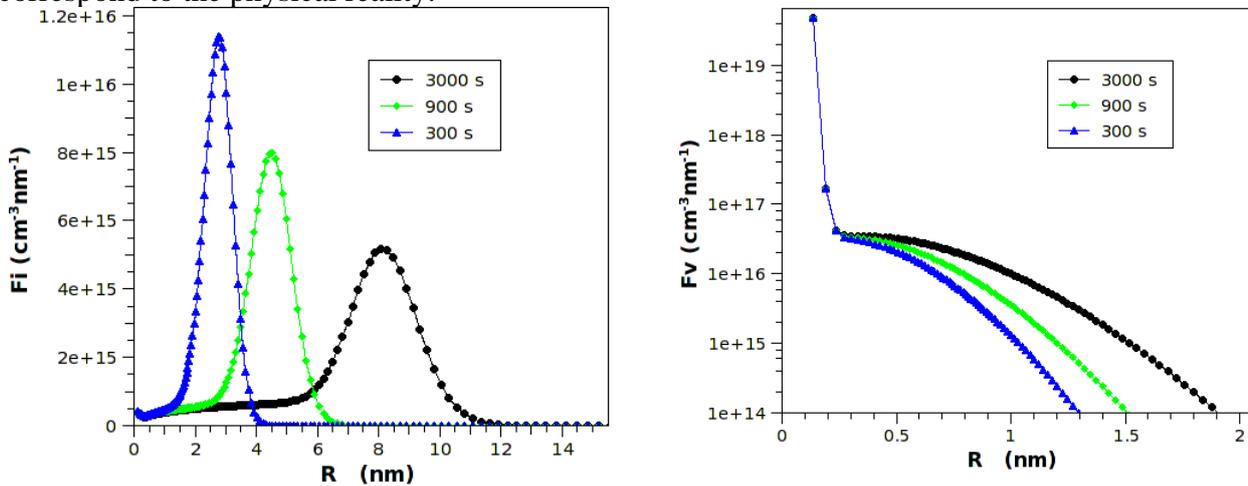

Figure 3: Evolution of the size histograms of (left) interstitial loops and (right) vacancy loops.

**Continuous isothermal irradiation.**

An example of the calculate evolution of the distribution of point defect clusters in the FeCu 0.1 at%, during continuous irradiation with 1 MeV electrons, at a damage rate of $1.5 \ 10^{-4}$ dpa/s is given on figure 3 [11]. In agreement with transmission electron microscopy experimental observations, interstitial loops nucleate and grow quickly up to large sizes when vacancy clusters remain under the observation limits.

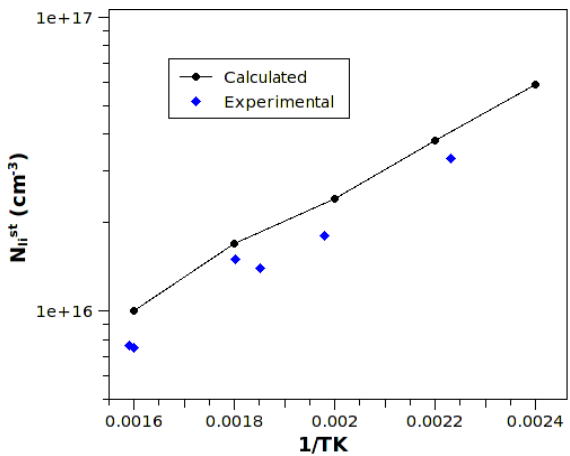

Figure 4: Evolution of the stationary number density of interstitial loops of radius larger than 1nm in normal iron.

The parameters needed in such simulations are numerous. All are not are not known. Adjustments cannot be avoided. The parameters used to fit the experimental results, as for instance the stationary total number of precipitates, are, as far as possible, closed to those calculated by atomistic methods (Fig.4). However, the effective vacancy migration energy used is approximately twice the calculated value in ultra pure iron to take into account the effect of the 100atppm carbon presents in the alloys [14]. Furthermore, it must be assumed that SIA clusters but single SIAs are immobile, in contradiction with the data obtained by atomistic calculations in ultra pure iron. This apparent contradiction could come to the trapping of SIA clusters by carbon atoms [14] or very small carbon clusters.

**On the effect of spatial correlations.**

In a particular case, a multi scale modeling has been performed using almost only parameters obtained from first principles. It is the simulation of the recovery of, at 4K, electron irradiated ultra pure iron, during isochronal annealing. The formation and migration energies of defect were obtained by using an ab initio methods and the kinetics, the Event Kinetic Monte Carlo (EKMC) code called JERK [15]. The various stages corresponding to the triggering of the mobility of one or more defects were correctly reproduced [16]. A comparison between CD and EKMC were carried out to check if the spatial correlations that appear in the defect creation process, have an important impact on the results [17]. Indeed, even with electron irradiation that does not induce displacement cascades, but isolated FP, vacancies and SIAs are not created at random: the distance between vacancy and SIA of the same FP, typically of four lattice parameters, is short in comparison with de mean distance between the centers of gravity of FPs.

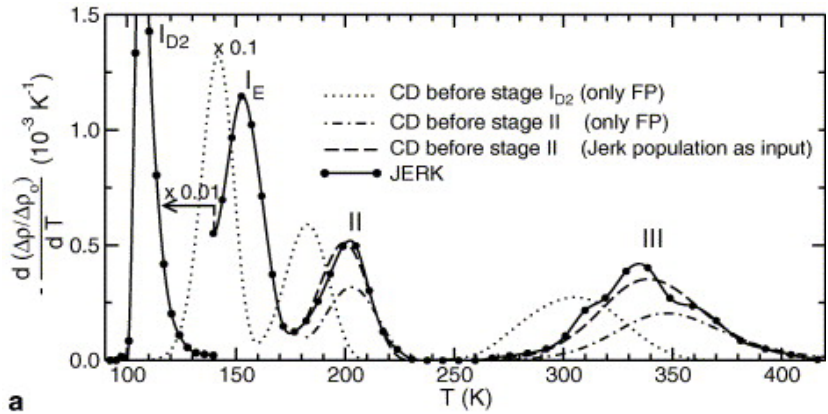

**Figure 5:** Comparison of isochronal recovery of Fe electron irradiated at 4K obtained with CD and EKMC.

As shown in the figure 5, CD gives good results if the initial conditions are not the total number of FPs after irradiation but what is given by KMC by the end of the stage I, corresponding to the recombination of PFs (mainly correlated). Indeed, if the initial point defect concentration is directly introduced at the beginning of the CD simulation, (i) the $I_{D2}$ and $I_E$ are merged together as the first one is due to correlated FPs and, (ii) the intensity of the first peak given by CD is smaller with the consequence that more point defects being present by the end of stage I, the intensity of higher temperature peaks are ten times too high.

This effect is less pronounced for higher initial concentrations of FP as the ratio of the distance between FPs to the distance between the vacancy and the SIA, of the same FP, decreases.

The difference between full KMC and full CD simulations would be more pronounced for irradiation with particles inducing displacement cascades such as neutrons or ions. As for electron irradiations, the source term used in CD should not be the one given by the end of molecular dynamics (MD) simulations but the one obtained after a short aging of MD cascades by KMC. The determination of the correct aging time to erase, as much as possible, the spatial correlations is not obvious. It is an important issue of the multi scale modeling using CD for kinetics. Another key point is the mobility of SIA clusters in actual steels that are certainly less mobile than what is given by atomistic simulations in ultra-pure material. The reason is that they are trapped by impurities and alloying elements.

It is worth noticing that the problem of correlations is less important for homogeneous precipitation as far as the precipitate volume fraction is small. Indeed the solute distribution at the beginning of the aging is very often random. It is also the case of the annealing of microstructures produced by continuous irradiations up to high fluences, which are currently random. It has been shown recently that CD can reproduced correctly the microstructure annealing of irradiated zirconium alloys [18].

**Conclusion**

The basic aspect of the Cluster Dynamic modeling (CD) and some applications in relationship with experiments are presented in this short paper. CD modeling is, in many cases, the only way to simulate long term evolutions of the microstructures of material submitted to thermal treatments or irradiation. The main advantage of CD is its low computational cost allowing short time calculation. Such a possibility is obtained thanks to an important drawback that is the loss of space correlations of the elements of the microstructures. This aspect must be considered particularly for irradiation under cascade conditions. The only way to overcome this problem is to age, during a short time, by a KMC method, the displacement cascades given Molecular Dynamics, in order to erase as far as possible the spatial correlations.